\documentclass[aps,prb,preprint,groupedaddress]{revtex4}
\usepackage{graphicx}

\begin{document}

%\preprint{APS/123-QED}

\newcommand{\tc}{La$_4$Ru$_2$O$_{10}$}
\newcommand{\htc}{ht-La$_4$Ru$_2$O$_{10}$}
\newcommand{\ltc}{lt-La$_4$Ru$_2$O$_{10}$}
\newcommand{\csro}{Ca$_x$Sr$_{1-x}$RuO$_3$}
\newcommand{\cro}{CaRuO$_3$}
\newcommand{\sro}{SrRuO$_3$}
\newcommand{\ssro}{Sr$_2$RuO$_4$}
\newcommand{\sssrro}{Sr$_3$Ru$_2$O$_7$}
\newcommand{\ccro}{Ca$_2$RuO$_4$}
\newcommand{\llco}{La$_2$CuO$_4$}
\newcommand{\ruoct}{RuO$_6$}
\setlength{\topmargin}{-0.5in}

\title{Properties evolution near the ferromagnetic quantum critical point in the series {Ca$_x$Sr$_{1-x}$RuO$_3$}}

\author{P. Khalifah}
\author{I. Ohkubo}
\author{H. Christen}
\author{D. Mandrus}
\affiliation{Oak Ridge National Laboratory, Condensed Matter
Sciences Division, Oak Ridge, TN 37831}

\date{\today}

\begin{abstract}

A series of epitaxial films were grown across the solid solution
\csro\ in order to pinpoint the ferromagnetic to paramagnetic
quantum phase transition in this system and to study the evolution
of transport and magnetic properties in its vicinity.  The
ferromagnetic $T_c$ of \sro\ was found to decrease linearly with
Ca doping levels up to 70\%.  Further doping resulted in the
abrupt elimination of ferromagnetic order, and the onset of low
temperature ($<$ 10K) non-Fermi liquid (NFL) resistivity of the
form $\rho \propto \rho_0 + AT^{1.5}$ for samples with $x \leq
0.75 \leq 1.0$.  The resistivity exponent of 1.5 matches that
previously observed for intermetallic alloys (such as MnSi) at
their ferromagnetic quantum critical points, indicating possible
universality of this NFL behavior.  Field-dependent specific heat
measurements on bulk samples at compositions near the quantum
phase transition provide additional evidence for NFL behavior
($C/T \propto \log T $) and show the conditions under which spin
fluctuations contribute to the specific heat.

\end{abstract}

\maketitle

%\setlength{\baselineskip}{20pt}
%\section{Introduction}

Ruthenates initially attracted the attention of the physics
community when superconductivity was discovered in \ssro\
\cite{maeno94}, a compound isostructural with the high-Tc
superconductor La$_{2-x}$Ba$_x$CuO$_4$ \cite{bednorz86}.  Since
that time, ruthenates have proved to be interesting in their own
right, with their 4$d$ magnetism
\cite{khalifah01,grigera01,mypaper6,nakatsuji00} generating as
much or more interest than their superconductivity. A focus area
of ruthenate research which draws on both their good conductivity
and their tunable magnetism has recently emerged, converging in
the study of non-Fermi liquid behavior in these systems
\cite{khalifah01,grigera01,klein99,capogna02}. The Fermi liquid
model of Landau predicts a standard behavior of metallic systems
at low temperatures, namely a resistivity $\rho$ which scales as
$T^2$, a temperature independent magnetic susceptibility $\chi$,
and a temperature normalized specific heat $C/T$ which is
constant.  The preconditions of this model ($T$-independent
electron interactions which are short range in both space and time
\cite{stewart01}) are almost universally satisfied among solid
state compounds.  The compounds which are the exception to this
rule (non-Fermi liquids, or NFL materials) are therefore of
interest. The study of NFL compounds had its genesis with
U$_x$Y$_{1-x}$Pd$_3$ \cite{seaman91}, and has now been extended to
a few dozen intermetallic systems \cite{stewart01}. NFL behavior
in oxides remains rare, and the best examples to date have been
found among ruthenates, most notably La$_4$Ru$_6$O$_{19}$
\cite{khalifah01}, \sssrro\ \cite{grigera01}, and \cro\
\cite{klein99}.

One potential route for finding NFL behavior is tuning a system to
the vicinity of a quantum phase transition (QPT), a regime where
the length scale of electronic fluctuations are diverging as the
precise end point of the quantum phase transition is approached.
This methodology has been used to study NFL behavior in systems
such as U$_x$Y$_{1-x}$Pd$_3$ \cite{seaman91} and
CeCu$_{6-x}$Au$_x$ \cite{lohneysen96a} whose antiferromagnetic
ordering temperature $T_N$ can be suppressed to zero by doping.
Additionally, NFL behavior can be found at a metamagnetic
transition, as exemplified on the field-tuning experiments on
\sssrro\ \cite{grigera01}. Similar behavior can be found by
suppressing the Curie temperature ($T_c$) of ferromagnets, as was
observed when hydrostatic pressure was applied to suppress the
weak ferromagnetism of MnSi \cite{pfleiderer97,pfleiderer01}.  Of
these three routes to QPTs, those involving ferromagnetic QPTs are
the rarest \cite{stewart01}, and the discovery and
characterization of additional systems with ferromagnetic QPTs
will provide essential information for the study of NFL quantum
criticality.

The perovskite \csro\ series is a good candidate system for
observing NFL behavior at the vicinity of a ferromagnetic quantum
critical point, given that \sro\ is a moderately strong itinerant
electron ferromagnet ($T_c$ = 160K) whose $T_c$ can be completely
suppressed via Ca substitution.  This system has been shown to be
metallic across its entire breadth \cite{eom92}.  Furthermore,
studies on epitaxial films of its end member \cro\ have found a
resistivity exponent of $\rho \propto \rho_0 + AT^{1.5}$
\cite{klein99,capogna02}. Specific heat studies on powder samples
have found a diverging $C/T$ near a Ca content of x = 0.8
\cite{kiyama98}. As powder samples are unsuitable for measuring
resistivity exponents, and no method exists for synthesizing
precisely targeted compositions of \csro\ crystals, this study
examined the resistivity of thin film samples in the \csro\
series.  In addition, a select set of powder samples with a
composition near x=0.8 were prepared to study the field-dependent
behavior of the specific heat near the critical regime.

%The itinerant nature of the electrons in ruthenates has allowed
%the observation of anomalous resistivity behavior in the vicinity
%of magnetic transitions, making this ruthenates perhaps the best
%oxide system for investigating non-Fermi liquid behavior.

%\section{Experimental}

\csro\ films were grown epitaxially on a lanthanum aluminum oxide
(LAO) substrate using a two-target pulsed laser deposition
technique.  The two targets of commercially obtained \sro\ and
\cro\ (Praxair) were mounted on a rotating carousel, and by
varying the relative number of laser pulses fired onto each target
the stoichiometry of the film could be precisely controlled.  For
all mixed samples, the maximum number of pulses fired
consecutively on one target was always less than 10 (note that 25
pulses are typically necessary to deposit a full perovskite
monolayer). For film growth, the substrates were mounted on a
heated plate that was kept at 675 $^\circ$C in a background of 80
mTorr of oxygen. KrF (248nm) eximer radiation was used with an
energy density of 3.87 J/cm$^2$ at the target.  The thicknesses of
the films were measured by cross-sectional scanning electron
miscroscopy (SEM) analysis, and were approximately 2000 \AA\ in
height. Film compositions were monitored both by energy dispersive
x-ray analysis (EDX) and Rutherford backscattering (RBS).

Powder \csro\ samples for heat capacity measurements were prepared
from high purity starting reagents of RuO$_2$ (Alfa, 99.95\%),
SrCO$_3$ (Alfa, 99.995\%), and CaCO$_3$ (Alfa, 99.995\%).  Ground
mixtures of the stoichiometric ratios of the reagents (with ~1 mol
\% excess RuO$_2$) were initially calcined at 1100 $^\circ$C to
react ruthenium with the carbonates, minimizing the Ru volatility
in future heating steps.  The product of this reaction was phase
separated into Ca-rich and Sr-rich perovskite fractions, as seen
in powder x-ray diffraction patterns collected on a Scintag PAD-V
diffractometer.  Further grinding and heat treatments ($\sim$48h
total) at 1300 $^\circ$C were necessary to obtain a single
perovskite phase.  Isostatically pressed pellets (1/2'' dia, ~2
tons pressure) were then prepared and fired at 1300 $^\circ$C.
Small sections were cut from these dense pellets to use in heat
capacity measurements.

Sample resistivities were measured using 4-probe measurements on
bar samples with areas of approximately 1 $\times$ 4 mm$^2$. Pt
wire leads (1 mil) were attached using silver epoxy (Epotek,
H20E). A short ($\sim$1 hr) high temperature anneal (300-500
$^\circ$C) was necessary to reduce the contact resistance below 5
$\Omega$. The dc resistance and magnetoresistance measurements
were done in the 2-350K temperature range in a Quantum Design
Physical Property Measurement System (PPMS) using a sufficiently
low current (typically 0.5 to 5.0 mA) to avoid sample heating.
Magnetoresistance measurements were made on heating after
previously cooling samples in the absence of a magnetic field.
Residual resistivities were on the order of 50-150 $\mu\Omega$ cm
while room temperature resistivities fell between 150 and 400
$\mu\Omega$ cm.  Heat capacity measurements were performed in the
Quantum Design PPMS using dense plates of pressed powder samples
with typical masses of 30-45 mg cut from the center of pellets.
Magnetic measurements were performed in a commercial magnetometer
(Quantum Design MPMS).

%\section{Results and Discussion}

%%%%%%%%%%%%%%%%%%%%% Discuss Sample Purity %%%%%%%%%%%%%%%%%%%%%%%%%%

\begin{figure} \begin{center}
\includegraphics[height=7in]{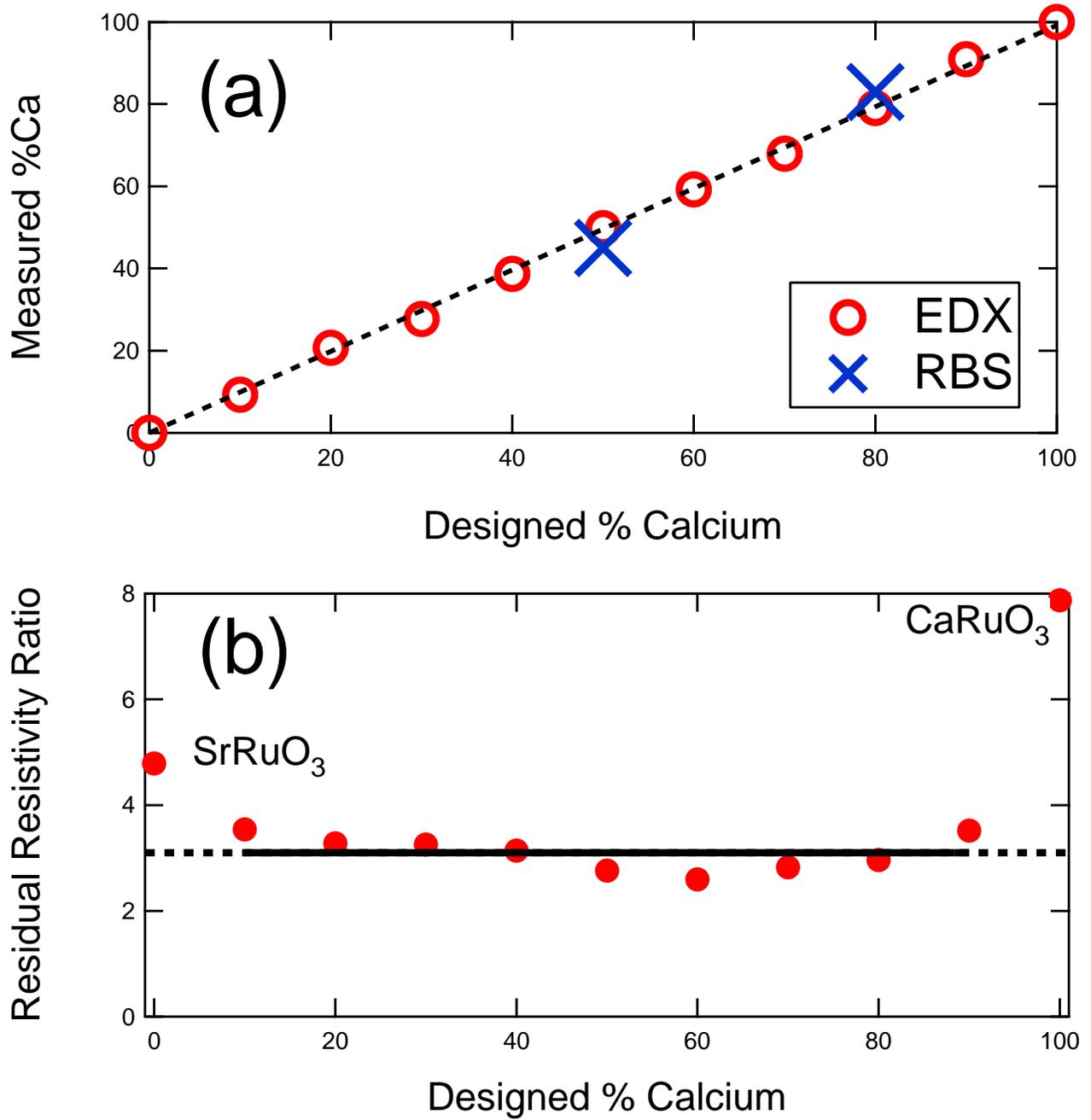}
\caption{Target and actual compositions of \csro\ films as
measured by elemental dispersive x-ray analysis (EDX, circles) and
Rutherford backscattering (RBS, crosses). (b) Residual Resistivity
Ratios (RRRs) of \csro\ films.} \label{COMP}
\end{center} \end{figure}

The primary advantage of thin film resistivity measurements over
single crystal measurements in the \csro\ solid solution is the
precise control of sample stoichiometries.  Due to the different
partition coefficients and different volatilities of \cro\ and
\sro, the growth of homogenous mixed \csro\ ruthenate perovskites
crystals presents a tremendous technical challenge. In contrast,
it is a facile process to accurately synthesize targeted
compositions of \csro\ thin films using our two-target PLD
process. In this study, the \csro\ composition was varied in 10\%
steps across the entire compositional regime, with one additional
sample containing 75\% Ca made to allow a finer mesh near the
hypothesized quantum phase transition in this system.  As can be
seen in Fig. \ref{COMP}a, the targeted and measured composition of
all the films were identical within the precision of the
measurement techniques.

An additional measure of film quality is the residual resistivity
ratio (RRR), here taken to be the resistivity at room temperature
divided by the resistivity at 4K.  It can be seen in Fig.
\ref{COMP}b that the \csro\ films typically have good RRRs of
$\sim 3$, with the \sro\ and \cro\ end members having extra high
RRRs of 5 and 8, respectively.  The lack of composition dependence
across the mixed \csro\ samples indicates that the reduced RRRs of
the quaternary oxides is a result of electronic inhomogeneity
(from the presence of two different types of alkali earth sites)
rather than strict compositional disorder.  A similar trend has
been observed in mixed \csro\ single crystals \cite{unpub2}.  The
RRRs of these samples measured here are comparable to those of
most previous preparations of \cro\ and \sro \cite{eom92,hyun02},
though we note that others have produced some exceptionally high
quality films by utilizing chemical vapor deposition (CVD)
techniques on strontium titanate (STO) substrates
\cite{kiyama98,mackenzie98,klein96b}.

%%%%%%%%%%%%%%%%%%%%%  Discuss ferromagnetism %%%%%%%%%%%%%%%%%%%%%%%%%%%%

\begin{figure} \begin{center}
\includegraphics[height=7in]{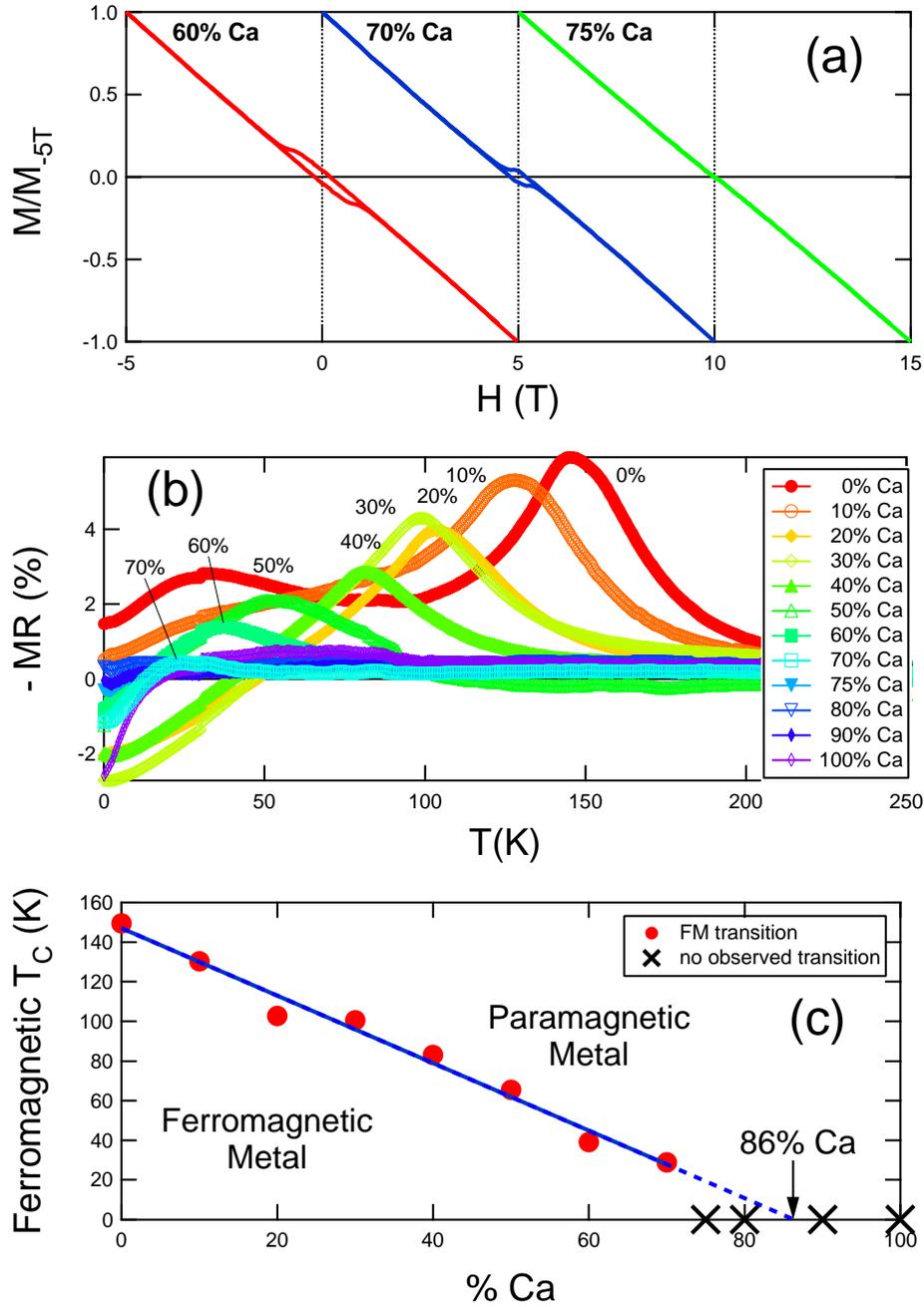}
\caption{(a) SQUID measurements of $x$=0.60,0.70, and 0.75. at T =
5K (b) Compositional dependence of magnetoresistance ($\rho_{0T}$
- $\rho_{8T}$). The break in the data sets at 30K is an artifact
of the slower ramp rate used below 30K. (c) Ferromagnetic phase
diagram of \csro\ films.} \label{TC}
\end{center} \end{figure}

Our first goal was to accurately determine the composition at
which the ferromagnetism of \sro\ is completely suppressed via
Ca-doping.  Magnetic susceptibility measurements (Fig. \ref{TC}a)
show that samples with 60\% and 70\% Ca have substantial magnetic
hysteresis, while the sample with 75\% Ca does not, placing the
endpoint of this quantum phase transition between $x$ = 0.70 and
$x$ = 0.75.  Due to the small mass of the deposited films ($\sim 1
\mu$g), it was impractical to determine the ferromagnetic ordering
temperatures from dc susceptibility measurements. Instead,
ordering temperatures were obtained from magnetoresistance (MR)
measurements, as shown in Fig. \ref{TC}b. Each ferromagnetic
sample has a well-defined peak.  Gaussian fits accurately
described the peak maxima, allowing a ferromagnetic $T_c$ to be
determined for each sample in the range 0 $\leq x \leq 0.70$. It
should be noted that the T$_c$ obtained from MR measurements is
analogous to $T_c$ obtained from ac magnetic susceptibility
measurements.  In both of these cases, the peak maximum occurs at
the midpoint of the transition where spin fluctuations are
largest.  This $T_c$ is always lower than the onset $T_c$ obtained
from dc magnetic susceptibility measurements.

The ferromagnetic phase diagram of \csro\ (Fig. \ref{TC}c) shows a
linear dependence of $T_c$ on composition down to 70\% Ca. If this
line is extrapolated to $T_c$ = 0, an endpoint is predicted at $x$
= 0.86.  This is in contrast to both the SQUID and MR
measurements, which show that the ferromagnetic ordering
disappears by $x$ = 0.75.  It appears that there is a precipitous
drop in the ferromagnetic $T_c$ in the vicinity of the \csro\
quantum phase transition.  The origin of this abrupt change is not
known, but will be the focus of future experiments using a finer
compositional grid near the transition temperature.

%%%%%%%%%%%%%%%%%%%%%% Discuss Resistivity %%%%%%%%%%%%%%%%%%%%%%%%%%%%%%%

\begin{figure} \begin{center}
\includegraphics[height=7in]{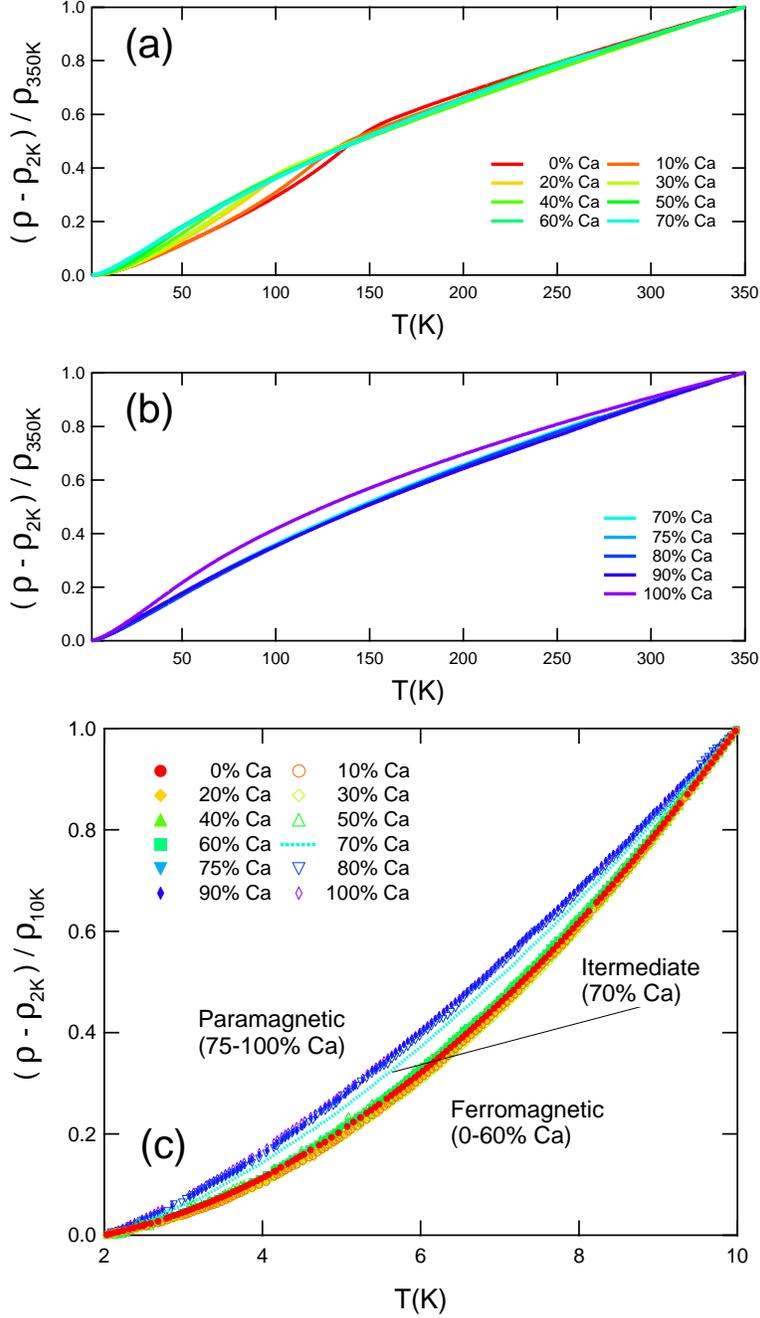}
\caption{(a) Normalized resistivities for 0-70\% Ca.  (b)
Normalized resistivities for 75-100\%Ca.  The resistance of the
60\% Ca sample (dotted line) is shown for comparison.  (c) Scaled
resistivities in the 2-10K range showing the different scalings of
the ferromagnetic and paramagnetic samples.} \label{RHO}
\end{center} \end{figure}

The effect of ferromagnetism on the resistivity of \csro\ can be
seen in Fig. \ref{RHO}a, which shows the normalized resistivity of
samples with 0-70\% Ca.  The high temperature resistivity data
show little compositional variation, though at lower temperatures
each of these ferromagnetic samples shows a marked reduction in
the resistivity on passing through the magnetic ordering
transition due to the reduced spin scattering in the ordered
phase.  As expected, the downturns in the resistivity curves show
monotonic compositional variation.

This compositional variation ceases for the nonmagnetic samples
(0.75 $\leq x \leq$ 0.90), indicating that there are no
significant differences in the spin scattering between these
compositions (Fig. \ref{RHO}b).  The failure of the 100\% Ca
sample to scale with the other nonmagnetic samples is a
consequence of the higher RRR of this sample rather than intrinsic
electronic differences. This can be seen in Fig. \ref{RHO}c, where
the data are renormalized to highlight the low temperature (2-10K)
variations of the resistivity, which reflect the the intrinsic
electronic interactions. The low temperature scaling of our \csro\
samples shows that the resistivities of the entire compositional
range collapse onto two primary curves.  The nonmagnetic samples
($0.75 \leq x \leq 1.00$) fall on the more slowly varying curve
and the ferromagnetic ($0.00 \leq x < 0.70$) ones fall on the more
rapidly varying curve. The only exception is the weakest
ferromagnet ($x = 0.70$), which falls on an intermediate curve.
This establishes the low temperature regime of 2-10K as the proper
regime for looking for scaling exponents and other quantum
critical phenomena.  It should be noted that this temperature
regime has also been shown to be the most significant one for
understanding the related ruthenates \sssrro\ \cite{grigera01} and
Ca$_{2-x}$Sr$_x$RuO$_4$ \cite{nakatsuji03}.

%%%%%%%%%%%%%%%%%%%%%% Discuss Exponents %%%%%%%%%%%%%%%%%%%%%%%%%%%%%%%

One hallmark of non-Fermi liquid behavior is power law scaling of
resistivity, of the general form $\rho = \rho_0 + AT^\alpha$,
where $\alpha$ is a number less than the normal value of 2
predicted by Landau theory.  In principle, the exponent $\alpha$
can give insights into the nature of a system, via comparison with
theoretical predictions \cite{stewart01}.  In the framework of the
spin fluctuation theories of Moriya \cite{Moriya95} and Lonzarich
\cite{lonzarich97}, resistivity is predicted to scale as $T^{5/3}$
in the vicinity of a three dimensional ferromagnetic transition.
This can be contrasted to the antiferromagnetic case, which is
instead predicted to show a $T^{3/2}$ dependence.  \cro\ was
previously reported to obey $T^{3/2}$ scaling in its resistivity
\cite{klein99,capogna02} over the range of 2-10K.

\begin{figure} \begin{center}
\includegraphics[height=7in]{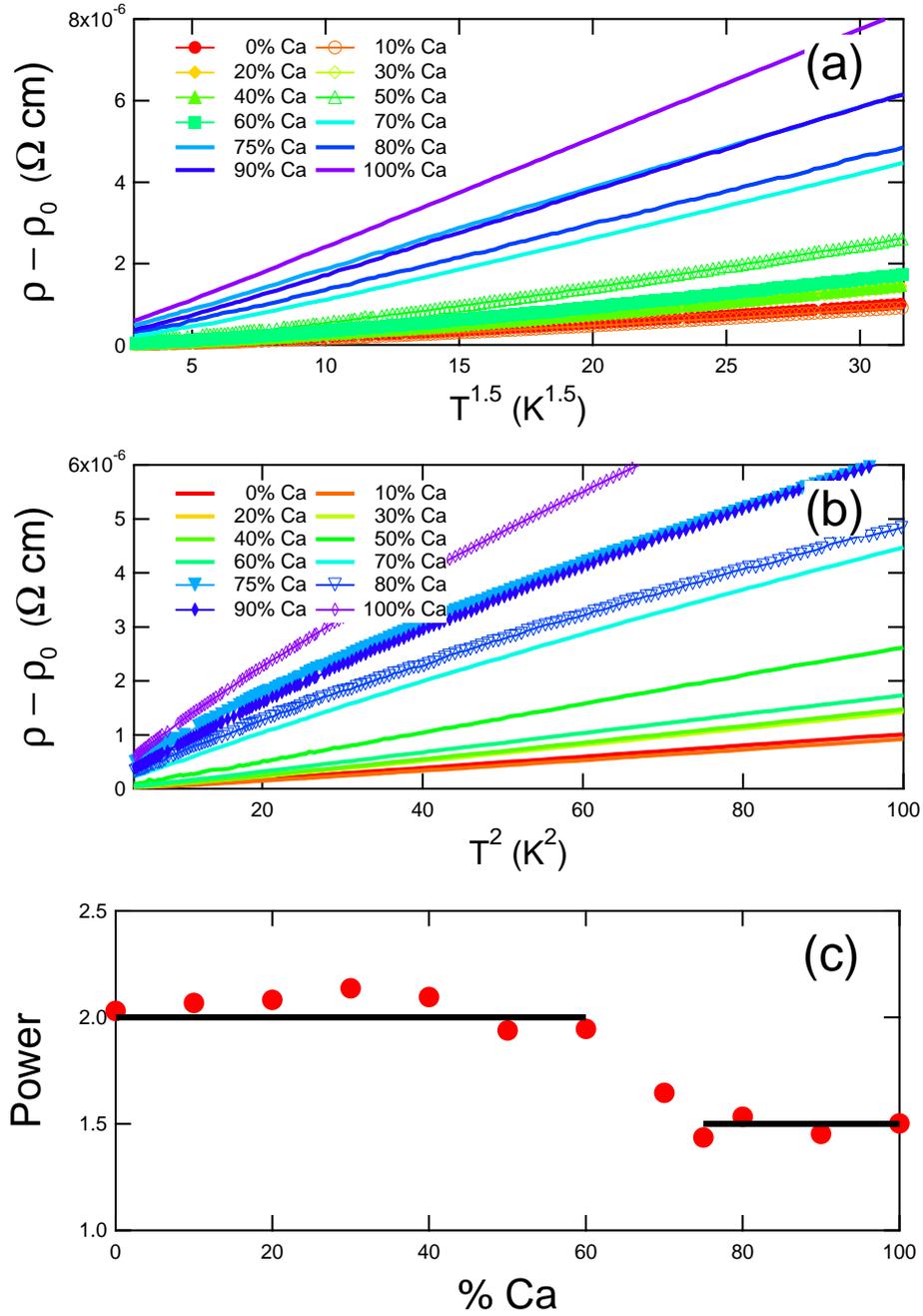}
\caption{(a) Intrinsic resistivity plotted as a function of $T^2$
(b) Intrinsic resistivity plotted as a function of $T^{1.5}$ (c)
Resistivity power ($\alpha$ in $\rho = \rho_0 +AT^\alpha$) as a
function of Ca concentration.  Solid lines are drawn at $\alpha$ =
1.5 and 2.0} \label{EXP}
\end{center} \end{figure}

Log-log plots of the normalized resistivity ($\rho - \rho_0$)
versus temperature showed nearly linear behavior below 10K over
our entire range of \csro\ samples, again confirming that the
search for power laws should be done in this regime (data not
shown).  As seen in Fig. \ref{EXP}a, normal Fermi liquid behavior
($T^2$ scaling) is observed for samples with $<$ 70\% Ca.  On the
other hand, samples with $\geq$ 75\% Ca are found to scale as
$T^{1.5}$ (fig \ref{EXP}b). The 70\% Ca sample is again found to
fall in an intermediate crossover regime, as seen when the
extracted resistivity power law exponents are plotted in fig
\ref{EXP}c. This convincingly demonstrates two important facts.
First, the nature of the conduction is fundamentally different in
the magnetic and nonmagnetic samples. Second, the scattering
processes must be fundamentally the same for each of the
individual samples within each class of samples (ferromagnetic and
paramagnetic). From this, we conclude that the most ferromagnetic
samples are behaving as Fermi liquids, while the \textit{entire}
range of nonmagnetic samples exhibit the same type of non-Fermi
liquid behavior originally reported for the end member, \cro. This
is important to note as disorder has often been invoked as a
necessary component of NFL behavior.  However, the same NFL
scaling of the resistivity data is observed both for the single
phase (\cro, $x = 1$) and solid solution (\csro, $0.75 \leq x <
1$) films studied here, indicating that crystalline disorder is of
minimal importance in these samples.

\begin{figure} \begin{center}
\includegraphics[height=7in]{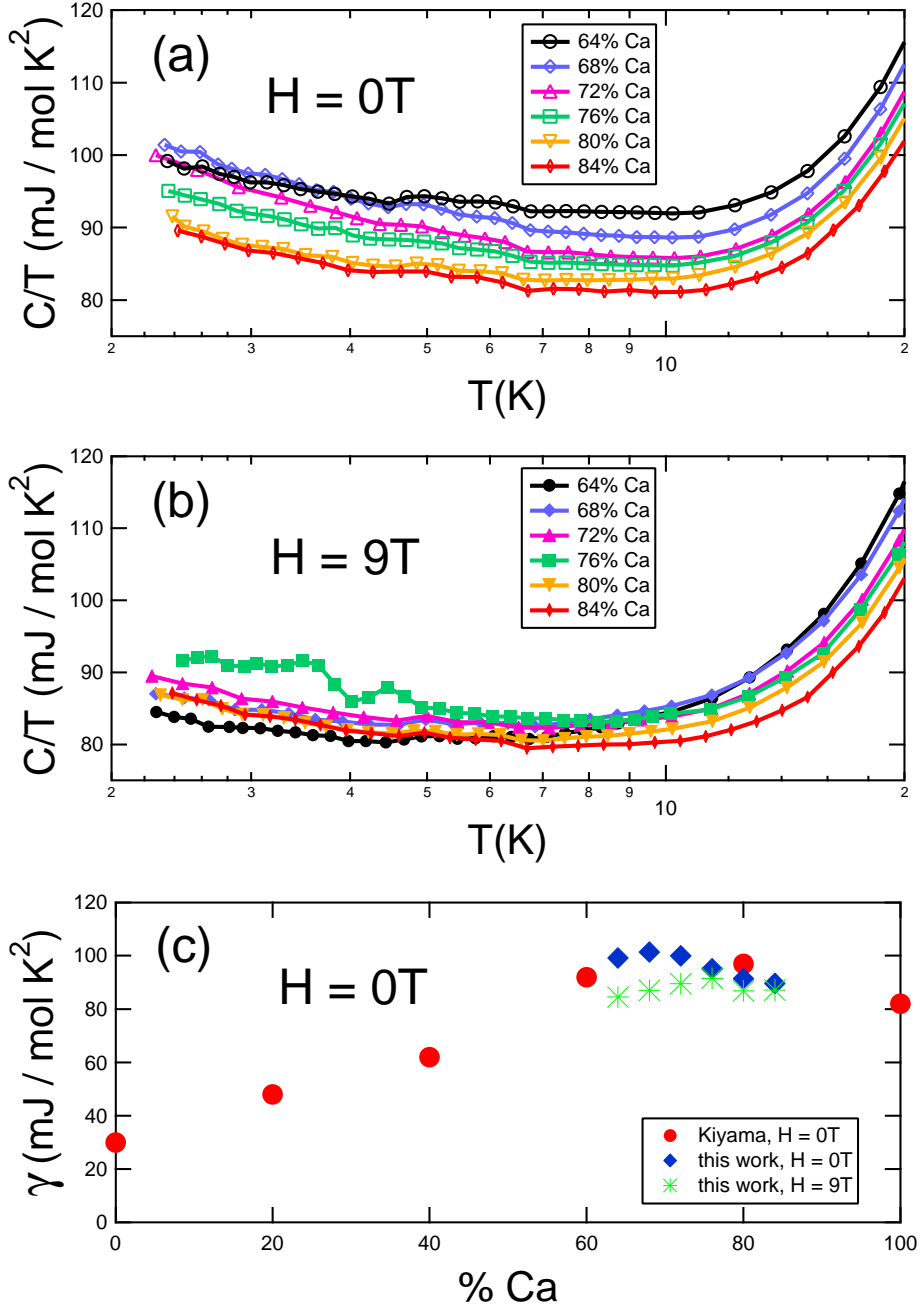}
\caption{(a) Zero field $C/T$ data, demonstrating $C/T \sim$
log$T$ below 7K. (b) High field (H = 9T) $C/T$ data.  Note the
generally lower magnitudes of $C/T$.  (c) Electronic
contributions, $\gamma$, to $C/T$.  In this work, $\gamma$ values
are taken from the measured $C/T$ at the lowest temperature
studied ($\sim$2K).  The values reported by Kiyama \textit{et al.}
were obtained by extrapolations to $T = 0$, with the different
methods potentially causing the discrepancy in the 80 \% Ca data.}
\label{HCHL}
\end{center} \end{figure}

\begin{figure} \begin{center}
\includegraphics[width=6in]{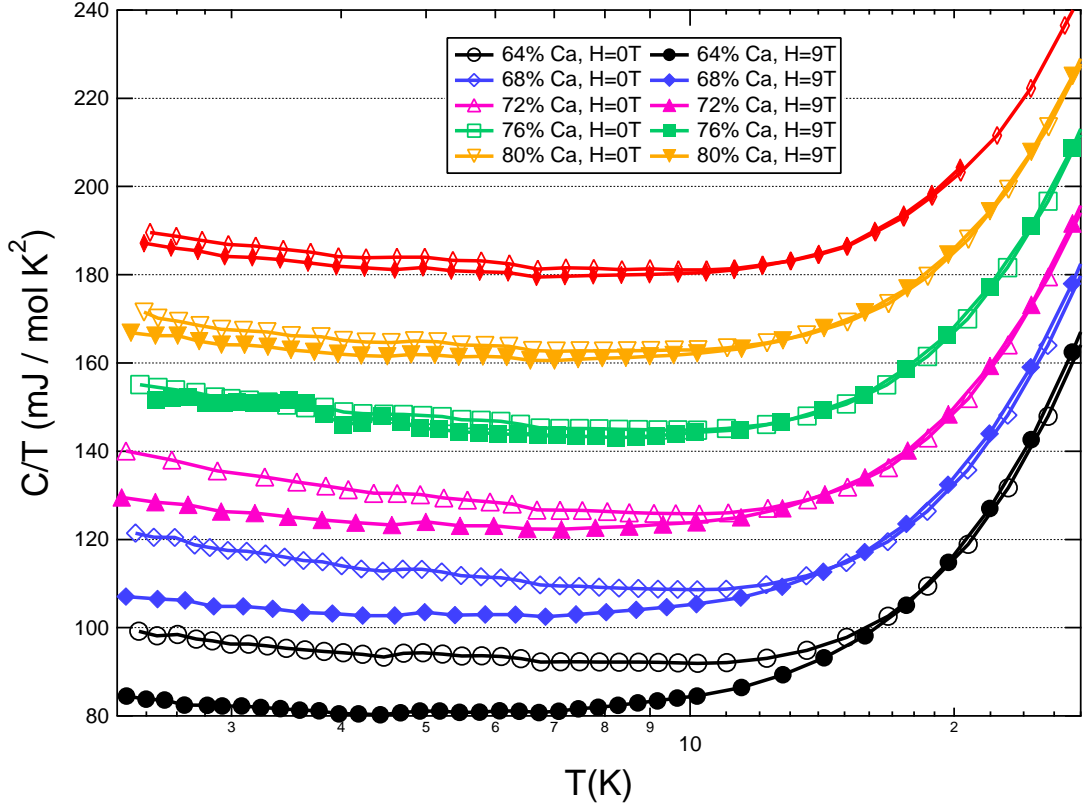}
\caption{Comparison of the zero field (open symbols) and high
field (solid symbols) specific heats of compositions near the
quantum phase transition in \csro, demonstrating the significant
spin fluctuation effects for 64-72\% Ca. Successive Ca
concentrations greater than 64\% Ca are offset by an extra 20
mJ/mol K$^2$ per step.} \label{HCCOMP}
\end{center} \end{figure}

Compounds exhibiting Fermi liquid behavior are expected to have a
heat capacity of $C$ = $aT^3 + \gamma T$ due to the phononic and
electronic contributions, respectively.  The zero field (H = 0T)
and high field (H = 9T) specific heat data for a range of
compositions near the quantum phase transition ($x$ = 0.64, 0.72,
0.76, 0.80, and 0.84) are shown in fig. \ref{HCHL}. At low
temperatures where the phononic contribution is negligible, Fermi
liquid theory predicts $C/T$ to have a constant value. This is not
observed in the \csro\ samples, which instead have a minimum in
$C/T$ plots around 10K.   As temperatures are decreased below 10K,
$C/T$ continues to increase with no apparent trend toward
saturation. This non-Fermi liquid behavior is most prevalent near
the ferromagnetic quantum phase transition in the \csro\ system,
as seen in fig. \ref{HCHL}a and also in previous measurements
\cite{kiyama98}. The $\gamma$ values exhibits a maximum near this
QPT (fig. \ref{HCHL}c), with the $\gamma$ for $x$ = 0.72 exceeding
100 mJ/mol K$^2$, more than three times higher than that of pure
SrRuO$_3$.  As seen in figs. \ref{HCHL}b and \ref{HCCOMP}, the
application of strong magnetic fields suppresses the electronic
portion of the specific heat by 10-15 mJ/mol K$^2$ for samples
with $x$ = 0.64 to 0.72, indicating that spin fluctuations play an
important role at these concentrations. These spin fluctuations
are only contributing near the QPT, as both samples with higher Ca
contents (fig. \ref{HCCOMP}) and pure \sro\ \cite{kiyama98} have
heat capacities which are field-independent.

Specific heat data provide additional evidence for quantum
critical scaling at the quantum phase transition in the \csro\
system.  One of the recognized hallmarks of quantum criticality is
a log $T$ dependence of $C/T$ \cite{stewart01,lohneysen96a}. In
plots of the zero-field (H=0T) specific heats versus log $T$ (fig.
\ref{HCHL}a), the entire range of \csro\ samples studied exhibit
linearity up to approximately 7K.  The compositions nearest the
QPT ($x$ = 0.68, 0.72) have the strongest temperature dependence
in addition to having largest low temperature $\gamma$ values.
This linearity is muted for the high-field data (fig.
\ref{HCHL}b), where the upturns in $C/T$ below 10K result in an
increase of only $\sim$ 5 mJ/mol K$^2$.

In conclusion, both resistivity and specific heat data indicates
that the non-Fermi liquid behavior of \cro\ results from its
proximity to the quantum phase transition where the ferromagnetic
moment of \sro\ is suppressed.  The NFL behavior which was
initially reported for \cro\ is not unique to that composition,
but is instead a universal feature of the \csro\ phase space with
0.75 $< x <$ 1.   The NFL resistivity exponent of $T^{1.5}$ found
for \csro\ samples is the same one observed when pressure is used
to suppress the itinerant electron ferromagnetism of MnSi, and has
also been reported for two other itinerant ferromagnets, ZrZn$_2$
and Ni$_3$Al \cite{pfleiderer01}. Given the widely disparate
nature of the \csro\ films and the MnSi single crystals (oxide vs.
intermetallic compositions, epitaxial films vs. single crystals,
residual resistivities of a few hundreds of vs. a few tenths of
$\mu \Omega$ cm, doping levels of up to 25\% vs. a pure system,
strong vs. weak crystalline electrical field effects), it is
intriguing that such similar behaviors are observed, and is
indicative of the dominant role that the nearly  magnetic
itinerant electrons play in determining the behavior of these
systems.

%Although it is often possible to reasonably fit the data in this
%range to a single power law, we have attempted to find the
%resistivity exponent as a function of temperature to more
%accurately follow the behavior within these samples.  A single
%residual resistivity ($\rho_0$) was obtained for each composition
%by a power law fit to the data in the low temperature regime where
%the power law behavior was best obeyed ($<$ 5K for ferromagnetic
%samples and $<$ 10K for paramagnetic samples).  The data were then
%fit over a sliding window of 25 points (a $\sim$1.4K range), with
%the center of the window being increased in $\sim$ 0.5K steps. The
%resulting plot of the power law as a function of temperature is
%shown in Fig. \ref{EXP}.

%\section*{Acknowledgements}

The helpful advice on experimental techniques from B. Sales, R.
Jin, and J. Thompson was greatly appreciated, as was the
theoretical guidance of A. Millis, D. Singh, and R. Osborn.
Special thanks go to V. Keppens and G. Petculescu for making
equipment at the University of Mississippi available for
exploratory measurements.  Research sponsored by the U.S.
Department of Energy under contract DE-AC05-00OR22725 with the Oak
Ridge National Laboratory, managed by UT-Battelle, LLC.

%In the case of quantum critical systems, where this power law
%behavior is though to arise as a result of the electron scattering
%by spin fluctuation

%In the theory of quantum critical systems (reviewed by Stewart
%(cite)), the scaling exponent is a good indicator of the
%underlying dimensionality of

%One well known weakly ferromagnetic system which can be tuned
%through a quantum phase transition is MnSi.  This ferromagnetic
%metal has a Curie temperature of about 30K, which can be
%suppressed to zero via the application of hydrostatic pressure
%(cite Pfleider). It has been found that the resistivity of MnSi
%varies as T$^1.5$ over almost three decades of temperature, from
%about 0.1 to 10K. A similar power law exponent has been previously
%been reported for CaRuO3 (cite Klein, Mackenzie), though
%milliKelvin measurements have not yet been reported for this
%phase.  In the phase diagram given for MnSi,

\newpage

\bibliographystyle{apsrev}

%\bibliography{structural,4210,CSRO,4619}

\vspace{3in}

%note -- see FeSi data from Fisk

\end{document}